\documentclass[epj]{svjour}
\usepackage{graphicx}
\begin{document}
\title{Fracture energy of gels}
\author{Yoshimi Tanaka\inst{1} \and Koji Fukao\inst{2}%
\and Yoshihisa Miyamoto\inst{2}
}                     
%
%
\institute{Department of Mechanical Systems Engineering, Faculty of
Engineering, Toyama Prefectural University, Kosugi-machi, 
Toyama 939-0398, Japan  \and
Department of Fundamental Sciences, Faculty of 
Integrated Human Studies, Kyoto University, Kyoto 606-8501, Japan}
\date{Received: date / Revised version: date}
%
\abstract{
To clarify effects of crack speed
and cross-link density on fracture energy of acrylamide gels, 
we evaluated the roughness of the fracture surface
and measured the fracture energy taking into account the roughness.
The fracture energy increases linearly with crack speed $V$
in a fast crack speed region, and the increasing rate of 
fracture energy with $V$ 
decreases with increasing cross-link density in the gels.
In a slow crack speed region the fracture energy
depends on crack speed more strongly than in the fast crack speed region.
This indicates that a qualitative change exists
in fracture process of the gels.
} 
\maketitle
\section{Introduction}
\label{intro}
Fracture is an old subject in material science. However, 
it is still extensively studied from the viewpoint both of
macroscopic mechanics (fracture mechanics)
and of molecular scale physics.
One of the recent topics in macroscopic studies
is an instability of fast cracks.
Experiments~\cite{Fineberg} and
simulations~\cite{Abraham} 
show that the steady state propagation
of a straight crack becomes unstable at a critical crack speed
of the order of the Rayleigh speed
in the material and accordingly roughening
and branching of the crack path occur. 
It is believed that
the instability can be essentially caused by the qualitative change
of the stress field which occurs as crack speed
approaches to the speed of surface wave~\cite{Langer,Freund}.
On the other hand, molecular scale processes
in vicinity of the crack front can also be
responsible for macroscopic behavior of fracture.
For example some metals undergo
the so-called brittle-ductile transition.
This transition is attributed to temperature dependence
of mobility of dislocations emitted on the crack front~\cite{Serbena}.
Effects of molecular scale process are quite different depending on the 
system in question and thus the unified understanding of fracture
is difficult even in the phenomenological level.

Systems classified in soft matter
often show elastic or viscoelastic nature
under normal conditions, and they undergo fracture
with well-defined crack front lines.
Fracture in the soft matter is interesting in the following
two aspects.
Firstly their structural units
have large spatial sizes
and show slow response against external forces.
Thus molecular process near crack fronts can drastically change
in slow fracture, which is experimentally controllable.
Secondly bulk behavior of the materials against external forces
deviates from the linear elasticity. To describe the fracture phenomena 
of the soft matter, we should extend usual
linear fracture mechanics
according to physical nature of the system
in question, for example, elastic large deformabiliy
of rubbers~\cite{Thomas,Andrews}, bulk viscoelastic effect
of polymer melt~\cite{DG96} and the anisotropic elastic energy
of smectics~\cite{DG90liq}.

Polymer gels~\cite{DG} are one of the typical systems
of the soft matter.
Various phenomena in gels such as gelation,
deformation and phase transition have been extensively studied.
Nevertheless, few studies have been carried out
on fracture of gels~\cite{Bonn,Tanaka96,Tanaka98}.
This is partially because gels have less industrial significance 
than 
hard solids do from the viewpoint of strength.
However, in the field of polymer physics,
this topic is interesting because we have much knowledge
on fracture processes in other polymer systems~\cite{Kinloch}
and on the various physical properties of polymer gels~\cite{Li},
which may be useful as a reference frame
to study fracture of gels.
For example, Gent and co-workers~\cite{Gent81} showed
that the strength of the butadiene rubber and
that of chemically-bonded interfaces
between the rubbers decrease with increasing cross-link
density of the rubber.
de Gennes~\cite{DG90} interpreted the effects as a consequence of 
friction during chain pulling-out process
on the crack front.
In gels the friction should be very small because they contain
large amounts of solvent, and it is not clear whether the frictional
dissipation is a dominant factor or not. 
The existence of solvent causes other unique effects,
such as the coupling between deformation of polymer
network and chemical potential of solvent.
To understand  fracture of gels,
we should first know experimental results which show fundamental
features of the phenomena.

In this study we investigate dependence
of fracture energy of acrylamide gel
on crack speed $V$ and on cross-link density
in order to get the fundamental experimental results.
Generally the fracture energy is determined by
microscopic process near 
crack fronts and appears in macroscopic
descriptions of fracture as an 
important parameter.
Thus, the fracture energy is an essential physical quantity to understand the nature
of fracture in gels. However, it is difficult to apply usual techniques for measuring 
fracture energy of hard solids to gels
because of their extreme softness and large deformability.
We have developed a novel method suitable
for measuring fracture energy of gels and precisely
measured the fracture energy of four kinds of acrylamide gel
with different cross-link densities.
In our method, we have adopted a peel test-like geometry
to drive fracture steadily and taken account of 
roughness of fracture surfaces to evaluate
the fracture energy of the gels.
Following results are obtained.
(i) For each cross-link density, the fracture energy
is an increasing function of $ V$
and when we change $ V$, a crossover occurs;
in the faster $ V$ side of the crossover 
the fracture energy linearly increases with $ V$ and in the slower $ V$ side,
where roughening of the fracture surfaces
is remarkable and the correction about the area of the fracture surfaces played an important role, 
the fracture energy depends on $ V$ with larger increasing rate than
in the faster $ V$ side.
(ii) At a given value of $ V$, both the value of fracture energy
and the increasing rate of the fracture energy with $ V$
decrease with increasing the cross-link density. 

\section{Experiment}
\label{Exp}

{ \it Samples---}
We use as samples four kinds of acrylamide gels
which have same polymer
concentration and different cross-link densities.
The amount of each reagent for preparing acrylamide gels is shown 
in Table 1.
Acrylamide monomer (AA, $M_w$ =71.08 )
constitutes sub-chains
and methylenebisacrylamide (BIS, $M_w$ =154.17)
constitutes cross-links.
Ammonium persulphate (APS) is an initiator
and tetramethylethlylenediamine (TEMD) is an accelerator of
the radical polymerization of AA and BIS.
We will distinguish the samples
by the codes of 4BIS $\sim$ 10BIS
as shown in Table 1.
To make pillar-shaped gels (2cm$\times$1.8cm$\times$14cm),
pre-gel solutions produced according to Table 1 were poured
into containers in which molds were arranged and
left for 24 hours at 25 $^{\circ}$C.
\begin{table}[b]
\begin{center}
\begin{tabular}{c||ccccc|cc}\hline\hline
   Sample code & Water & AA   & BIS   & APS  & TEMD   & $E$(dyn/cm$^2$)    & $V_
t$(cm/s) \\\hline
         4BIS  & 100cc & 10g  & 0.04g & 0.1g & 0.25cc & 0.56$\times$10$^5$ & 135
         \\
         6BIS  & 100cc & 10g  & 0.06g & 0.1g & 0.25cc & 1.21$\times$10$^5$ & 200
         \\
         8BIS  & 100cc & 10g  & 0.08g & 0.1g & 0.25cc & 1.86$\times$10$^5$ & 248
         \\
         10BIS & 100cc & 10g  & 0.10g & 0.1g & 0.25cc & 2.77$\times$10$^5$ & 302
     \\\hline\hline
\end{tabular}
\caption[table]{The composition of each reagent required to prepare
the four kinds of acrylamide gels, Young's modulus $E$ and
velocity of transverse wave $V_{t} $ of the gels.}
\end{center}
\end{table}
The gels are took off from the molds
and used for the fracture experiment.
The values of Young's modulus $E$ and
velocity of transverse wave of the gels $V_{t}$
are also shown in Table 1. The values of $E$ are measured
by compressing the gels. To calculated $V_{t}$
we used the value of density $\rho$ = 1.01 (g/cm$^3$),
and the value of Poission's ratio $ \nu$ = 1/2.

{ \it Peel test like method---}
In order to measure the fracture energy of the gel
we developed a method which is similar to the peel test.
In Fig. 1 we present a gel fractured by the method.
The set-up of the experiment is as follows.
We put the pillar-shaped gel on an aluminum plate and
heated the aluminum plate with a gas burner for about ten seconds.
By this treatment the sample gel is fixed on the aluminum plate.
We attached a strip of a filter paper to the upper surface of the
gel.
The filter paper is tightly absorbed to the gel.
We created an initial notch on one of the smallest surface
of the gel and made it propagate by 2cm
by pulling the filter paper with hands.
An end of the filter paper was connected to a stepping motor
located well above the gel (1.8m)
through a strain-gauge and a wire.
A thin layer of the gel ($\sim$1mm) is peeled off by rolling up the wire
with the stepping motor.
The control parameter is the rolling speed $V$ which
is equal to the crack speed $V$ in our 90$^{\circ}$ peeling
geometry. The measured quantity is the peeling force $F(t)$.

{ \it Fracture energy---}
Fracture energy $ G$ is defined as the energy needed
to make a unit area of a fracture surface.
In our peel-test like method the fracture energy $ G$ is calculated by the
following
equation,
\begin{equation}
G = \frac{F}{w}
\end{equation}
where $ F $ is the measured force
and $ w $ is the width of the pillar-shaped gel (see Fig. 1).
In order to understand this relation, let us
suppose that the crack front in Fig. 1 steadily
propagates over a distance $ \Delta x$ along the direction
of the longest axis.
The increase in area of the fracture surface by this propagation 
of the crack front is $ w \times \Delta x$
and the energy required to extend the fracture surface is
the work done to the gel, $F \times \Delta x $.
Therefore the fracture energy $G$ is
$ (F \times \Delta x) /(w \times \Delta x) $.
This is identical to the quantity mentioned above.

{ \it Roughness of fracture surfaces---}
In this study we evaluate the roughness of the fracture surfaces
using replicas produced
by molding the fracture surfaces using silicon rubber.
The replica was cut along the plane which was normal 
to the global fracture surface
and parallel to the direction of fracture propagation.
The shape of the cross section was recorded by an image scanner 
(Microtek, ScanMaker III).
A quantity which can be regarded as a measure of the roughness of the 
fracture surfaces was extracted from the image of the cross section.
\newpage

\section{Results}
\label{Res}

Figure 2a is $F(t)$ at $V=$ 0.4cm/s.
The arrows indicate the initiation and
the termination of the fracture propagation.
The fracture propagates steadily 
in the period of time between the arrows.
We divided the period of time corresponding
to the steady state fracture propagation equally into three periods and
evaluated the fracture energy $G$
using the time average of $ F(t)$ for the central period.

Figure 2b is $ F(t)$ at $ V=$ 0.04 cm/s.
As shown in Fig. 2a there is a period of time
corresponding to a steady sate fracture propagation.
However the fluctuation of $F (t)$ in Fig. 2b is larger than that
in Fig. 2a even in the
period of the steady state fracture propagation.
The increase in fluctuation of $F (t)$ is characteristic
of slow fracture and is accompanied with roughening of fracture surfaces.
We show evidence for the existence of the roughening of fracture 
surface later (see Fig. 5 and Fig. 6).

Figure 3 is a plot of the fracture energy $ G $
as a function of crack speed $ V $.
At fast values of $ V$ ($V > $ 1cm/s),
$G(V)$ depends linearly on $ V $ and
both $G(V)$ and $ dG/dV $ decrease with increasing
BIS concentration of the samples.
Figure 4 is a plot of the fracture energy $ G (V)$
of 4BIS, 6BIS and 8BIS for $ V < $ 1cm/s.
A common feature of $G (V)$ for these samples is that
there is a region of $ V$ where $G$ increases with decreasing $V$,
and $G(V)$ has a minimum (shown by the upward arrows in Fig. 4) at
a value of $ V$. Hereafter, we will call this $V_{min}$.
The dependence of $ G(V)$ on BIS concentration is
non-monotonical below the value of $ V  \cong  $ 0.8cm/s.

As $ V$ decreases across $ V_{min}$,
the roughness of the fracture surfaces grows up
(the roughening at {\it slow fracture}).
In Figures 5a-c we show the morphologies of fracture surface
of 6BIS at different crack speeds. The bars represent 0.9~cm.
Figures 5e-g show the cross sections of the fracture surfaces
shown in Figs. 5a-c, respectively.
The cross section is along the plane
which is perpendicular to the global fracture surfaces
and contains the center lines of the fracture surfaces
(the $ x$-axis in Fig. 5d).
The vertical size of the cross section corresponds
to 3cm and the horizontal size is magnified
2.5 times compared with the true scale.
The shape of right-hand side boundary of the cross-section
corresponds to the $h(x)$ shown in the illustration,
i.e., the height of the fracture surface measured
at each point of the $ x$-axis.
Figure 5a is a fracture surface of 6BIS above $ V_{min}$.
At such crack speeds most parts of fracture surface
are flat and a few steps exist on the global fracture surface
which seem like lines in Fig. 5a.
Around $V_{min}$,
such steps are frequently produced and the roughness
of the fracture surfaces begins to grow up (Fig. 5b).
As $V$ decreases further,
the roughness of the fracture surfaces becomes more remarkable (Fig. 5c).

To quantify the roughness of the fracture surfaces we introduce a quantity
$R$ defined by the following equations. 
\begin{eqnarray}
 R \equiv  \int_{l_{c}} \sqrt{1+(dh/dx)^2}~dx \Bigg/ \int_{l_{c}} dx  \\
=\Bigl< \sqrt{1+(dh/dx)^2} \ \Bigr>,  \nonumber
\end{eqnarray}
where the range of integration  $ l_{c}$ represents
the distance along the $x$-axis which corresponds to the central period of time in which the average of $F(t)$ is took (see the first paragraph of this section),
and the symbol $ \bigl<\cdots \bigr> $ represents the spatial average over the distance. The numerator on right-hand side in (2)
is the contour length of $h(x)$ over the distance,
thus $R$ is equal to 1 for the completely flat fracture surface ($ dh/dx = 0 $)
and increases from 1 as the roughness of the fracture surface increases.
Therefore, $R$ is an index of the roughness of the fracture surfaces.

In Fig. 6 we show  $ R $ as a function of the crack speed $ V $
for four kinds of sample gels.
$R(V)$ of the gels has a common feature; i.e.
at fast values of $V$, $R$ is close to 1 and with decreasing $V$,
$R$ begins to increase at the value of $V$
close to $ V_{min}$.
This fact clearly shows the correlation
between the roughening of fracture surfaces
and the increase in $G(V)$ with decreasing $ V$
across $ V_{min}$.

When we take into account the roughness of fracture surfaces,
we should correct the fracture energy by dividing it by $R^2$.
Strictly speaking, we should measure the same quantity as $ R$
along the lateral direction in Fig. 5, $R'$,
and divide $G(V)$ by $RR'$. However, the structures causing the roughening
are the steps on the fracture surfaces extending at 45$^{\circ}$ from
the $x$-axis. Therefore we can expect $R$ and $R'$ are very close.
In Fig. 7 and Fig. 8, we show the corrected
fracture energy $ \overline{G}(V) \equiv G(V)/R(V)^{2}$.
Behavior of $ \overline{G}(V)$ at fast values of $ V$ is
qualitatively identical to that of $G(V)$,
i.e., $ \overline{G}(V)$ linearly increases with $ V$
and $ \overline{G}(V)$ and $ d \overline{G}/ dV$
decrease with BIS concentration.
On the other hand, when $ V$ decreases, 
the crossover in $ \overline{G}(V) $ occurs in the narrow range
of $ V$,
and below the crossover range
$ d \overline{G} /dV$ becomes larger than above the crossover range.
As a result of the correction, $ \overline{G}(V)$
at each value of $ V$ in the region
monotonically depends on BIS concentration as in the fast $V$ region.

Our results for the corrected fracture energy $ \overline{G}$ can be summarized as follows:
\begin{enumerate}
\item[i)] At a given value of $V$, $ \overline{G}(V)$ decreases
with increasing BIS concentration.
\item[ii)] At fast values of $V$ ($V >$ 1cm/s),
$\overline{G}(V)$ for each BIS concentration
linearly increases with $V$.
\item[iii)] The increasing rate $ d \overline{G}/dV$
decreases with increasing
BIS concentration.
\item[iv)] With decreasing $ V$ across a
crossover range,
$ d \overline{G}/dV$ becomes larger.
\end{enumerate}

\section{Discussion}
\label{Dis}

In the first half of this section we will discuss
the corrected fracture energy $ \overline{G}$.
As shown in Figs. 7-8, the order of the fracture energy 
$ \overline{G}(V)$ of the gels is several hundred
times as large as that of the surface tension of water
(about 72 dyn/cm at 25$^{\circ}$C~\cite{Handbook}). 
Thus $ \overline{G}(V)$ reflects
energy needed for breaking the network structure of the gels
near crack fronts
and consists of two parts;
one is due to cutting polymer chains of the gel network $G_{cut}$,
the other is due to viscous resistance $G_{vis}$.
\begin{equation}
\overline{G}=G_{cut} + G_{vis}.
\end{equation}

Gels synthesized from monomer solutions contain various kinds of defects.
Characterization of the gels is open problems
in polymer physics; We can not give quantitative discussions
about our results from the microscopic viewpoint. 
However, the elastic modulus of the gels used in this study
increases with BIS concentration as shown in Table 1.
This shows that actual cross-link density increases
with BIS concentration.
The following qualitative discussion can be made on $ \overline{G}(V)$.

We first consider the number of polymer chains
cut on the fracture surface of the gels.
If all cross-linkings of a gel disappear, the system would be
a solution of linear polymer,
and we could divide it into two pieces without cutting
any polymer chain.
On the threshold cross-link density of gelation, we need cut 
a finite number of the polymer chains.
With increasing the cross-linking density,
we need to cut more number of the polymer chains
on the fracture surfaces. From this consideration, 
we expect that {\it $G_{cut}$ increases with BIS concentration.}

We next consider the number of elements which cause viscous resistance
for extension of a fracture surface.
Dangling chains, which have free ends, should be such structures.
Another possibility is the sub-chains long enough to
penetrate into other part of the gel.
As the cross-linking density decreases,
i.e., as network structure of gels becomes looser,
the number of the elements increases.
Thus, we can expect that 
{\it $G_{vis}$ increases with decreasing the BIS concentration.}

From above consideration, we can conclude that the result i) shows $G_{vis}$
overwhelms $G_{cut}$ at the values of $V$ accessed in this study.
The result ii) shows that $G_{vis}$ can be represented
by the following form,
\begin{equation}
G_{vis} =  \alpha V .
\end{equation}
The result iii) shows that the prefactor $ \alpha $
decreases with increasing the BIS concentration. This is reasonable 
because $ \alpha $ should be an increasing function of
the number of the elements contributing to viscous resistance.

The physical meaning of iv) becomes clear if we exchange
the ordinate and the abscissa of Fig. 7
and we recall that $ \overline{G}$ is proportional
to the force driving the fracture.
Above a critical value of $V$,
an increase of the driving force $\overline{G}$
causes larger increase of $V$ than below the critical value.
This type of nonlinear relation between a driving force and
the conjugate rate
is often observed in soft polymeric systems; for example,
the relation in stress/strain rate
of polymer melts (shear thinning)~\cite{Doi-Ed}
and the relation in loading/detaching speed of glass-rubber interfaces
stitched by linear polymers~\cite{Brown}, etc.
In essence these phenomena are explained by conformational
change of polymers due to the driving force~\cite{Ajdari}.
This mechanism may be applicable to fracture of gels.

At present, we do not identify the origin of $ G_{vis}$.
A possibility is the bulk viscoelastic dissipation.
Moreover, strong nonlinear process localized near
crack fronts, for example, chain pulling-out, may participate
in $ G_{vis}$.
To clarify this point, we need another experiment
where we use more controlled gels.

Now we discuss the phenomena related to the roughening of
fracture surfaces.
As shown in Figs. 4-6, the apparent fracture energy $G(V)$
increases with decreasing $V$, accompanied 
with the roughening of the fracture surfaces.
This means that fracture of the gels does not follow the path
of minimum dissipation. In other words,
under slow fracture conditions gels increase their strength 
by undulating crack front lines.
On phenomenological level, similar $V$-$G$ curve is reported in fracture of
glassy polymers such as PMMA~\cite{Kinloch,Kausch}, which is attributed to crazing.
The crazing, which is a kind of plastic deformation, is characteristic of
glassy polymers, and we can not expect concrete relation in molecular level
between the $V$-$G$ curve of gels and that of glassy polymers.

Scale invariant nature of rough fracture surfaces is
one of the topics in physical study of fracture.
Theoretically, it has been studied as a stochastic
effect~\cite{Bouchaud}.
In the gels, the roughening is caused by the definite elements,
i.e., steps
on fracture surfaces extending to the direction of
$\pm$ 45$^{\circ}$ from the $x$-direction.\footnote{Besides the
oblique step-lines,
scratch-like steps extending along
the $x$-direction can be seen in Figs. 5a-c. From qualitative
observation by eyes and by low magnification microscope,
following tendency is recognized: (i) Large(thick) steps are
the oblique type and small(thin) steps are the scratch-like type. 
(ii) the critical height of steps becomes
larger as $V$ increases. 
Two kinds of morphological transition
occur on fracture surfaces of gels:
One is the roughening transition which is related to nucleation
frequency of
the steps on the whole of a crack front,
the other is the oblique/scratch-like
transition of each step-line. In a previous study~\cite{Tanaka96}
where we used acrylamide gels
of higher polymer concentration and
lower cross-link density than those used in the present study,
we confused these two transitions because
in the gel used in the previous study the coexistence of
the two kinds of steps occurs
only in relatively narrow range of $V$
($V$ $\sim $ 0.5~cm/s) and the roughening
transition occurs in the same range.
We will report on details of the morphological transition
elsewhere.}
(In a previous work~\cite{Tanaka98} we clarified the
structures on crack fronts which create the oblique
step-lines, and classified collision process
between the structures.) This result gives a new viewpoint
on the general study for the rough fracture surfaces.

Wallner~\cite{Wallner} reported similar
oblique step-lines on
fracture surfaces on glass
(Wallner lines). The proposed mechanism~\cite{Holloway} is that
the Wallner lines are created at the parts of a moving crack front
where stress field is disturbed by stress pulses
which are nucleated when the crack fornt
passed through irregular points of the matterial, i.e.,
the Wallner lines are loci of the intersections
between the crack front and fronts of stress pulses.
This mechanism does not hold for the step-lines of
gels because the $\pm$45$^{\circ}$oblique step-lines of gels
are observed even at much slower values of $V$
than the sound velocity in the gels;
in fracture process in Fig. 5c, for example,
the crack front ($V=$ 0.015cm/s) is almost stationary compared with
the stress pulses ($V_t$=200cm/s, see Table 1),
and the loci of the intersections
are almost along the crack front itself, i.e.,
almost horizontal in Fig. 5c.
This does not agree with the $\pm$45~$^{\circ}$
obliquity of the step-lines.
On the other hand, with regard to geometrical aspect 
it is probable that the Wallner lines are created by
similar structure on crack fronts to that observed in gels.

In summery, we studied dependence of the fracture 
energy on the crack speed $V$ and on cross-link density,
taking account of the roughness of the fracture surfaces. 
The following features is found on
$V$ dependence of the fracture energy;
(i)At a given value of $V$, the fracture energy decreases with the cross-link density.
(ii)At fast values of $V$ ($V>$ 1cm/s), the fracture energy
linearly increases with $V$,
and at slow values of $V$ 
the fracture energy increases with $V$ with larger increasing rate than
at fast values of $V$. 
These results indicate that the dissipative effcts dominate over 
the effects of the breakage of polymer chains in the fracture of gels.

The authors thank Professor Ken Sekimoto for conducting us to study of
this field.
They also thank to Professor Fumihiko Tanaka
and Professor Mitsugu Matsushita
for their helpful comments. 
This work was partly supported by a Grant-in-Aid from the Ministry of 
Education, Science, Sports and Culture of Japan.

\newpage


\begin{figure}
\begin{center}

\includegraphics*[width=8cm,keepaspectratio]{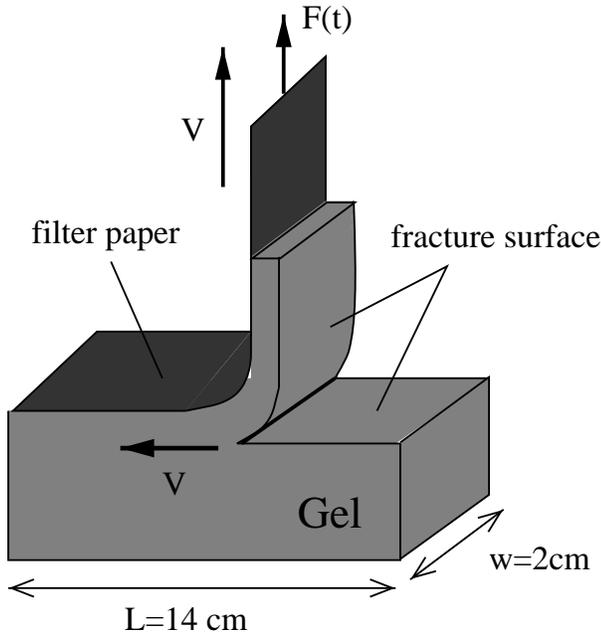}

\end{center}

\caption{A schematic of a gel undergoing fracture. We fixed the gel on
the aluminum plate and made an initial notch
and attached the filter paper to the upper surface.
By pulling-up the filter paper vertically at a constant speed $V$,
the crack propagates
through the gel at the rate $V$.
The force needed in pulling-up $F(t)$is measured with a strain-gauge.}
\label{fig:fig1}
\end{figure}


\begin{figure}
\vspace{2cm}\begin{picture}(360,1)
 \put(10,-40){(a)}
 \put(10,-140){(b)}
 \end{picture}
\begin{center}
\includegraphics*[width=7cm,height=6cm]{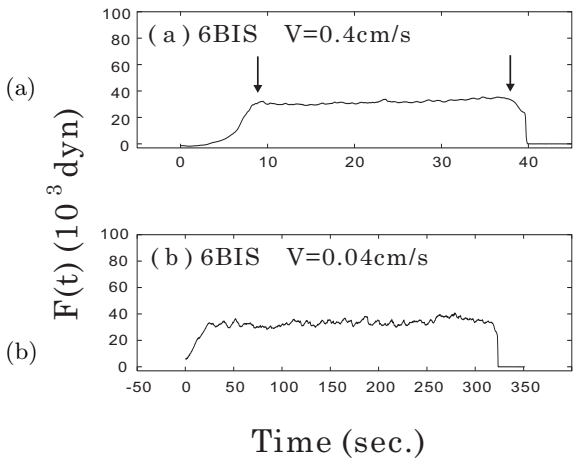}
\end{center}

\caption{Examples of measured peeling force $F(t)$.
(a) is an example of $F(t)$ at $ V= $ 0.4cm/s.
The arrows indicate the initiation and the termination of
fracture. In the period of time between the arrows,
the fracture steadily propagated.
The fracture energy $G$ was evaluated from the averages of $F(t)$
in the central part of the period.
(b) is an example of $F(t)$ at $V= $0.04cm/s.
Fluctuation of $F(t)$ is larger compared with (a).}
\label{fig:F_t}
\end{figure}



\begin{figure}
\begin{center}
\includegraphics*[width=8cm]{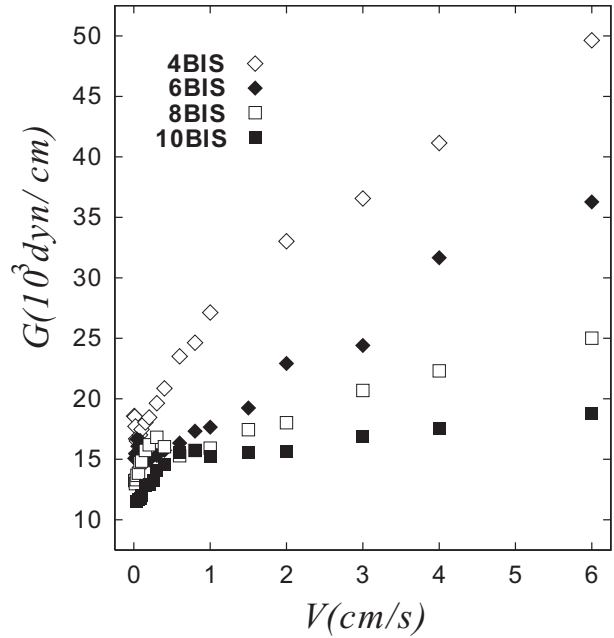}
\end{center}

\caption{The fracture energy $G(V)$.
At fast values of $V$ ($V>$ 1cm/s),
$G(V)$ of each sample linearly increases with $V$.
In the region, $G$ at a given value of $V$
decreases with the BIS concentration.
At slow values of $ V$,
$G$ depends on $V$ in more complex fashion.}
\label{fig:G_v}
\end{figure}

\begin{figure}
\begin{center}
\includegraphics*[width=7cm]{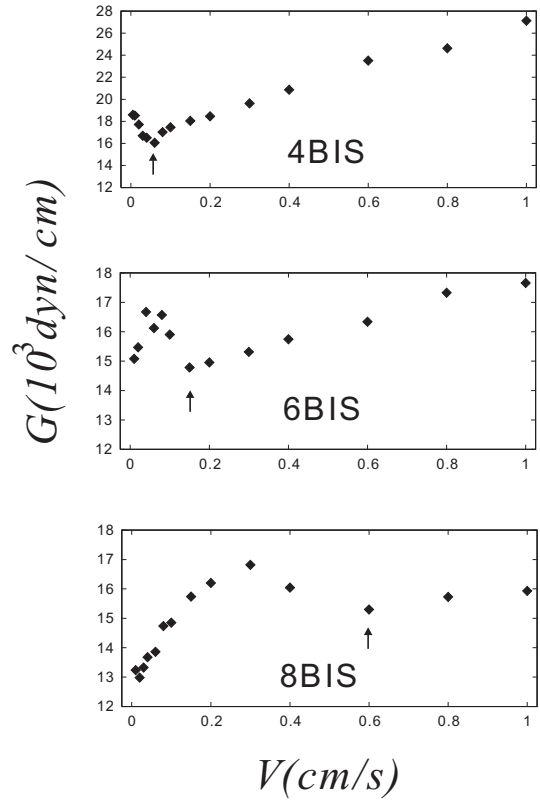}
\end{center}
\caption{$G(V)$ of 4BIS, 6BIS and 8BIS
for $V<$ 1cm/s.
$G(V)$ non-monotonically depends on $V$
in the region and the minimum of $G(V)$ exists.}

\label{fig:G_v2}
\end{figure}

\pagebreak
\begin{figure}
\begin{center}

\vspace{4cm}
{\large See attached jpg file (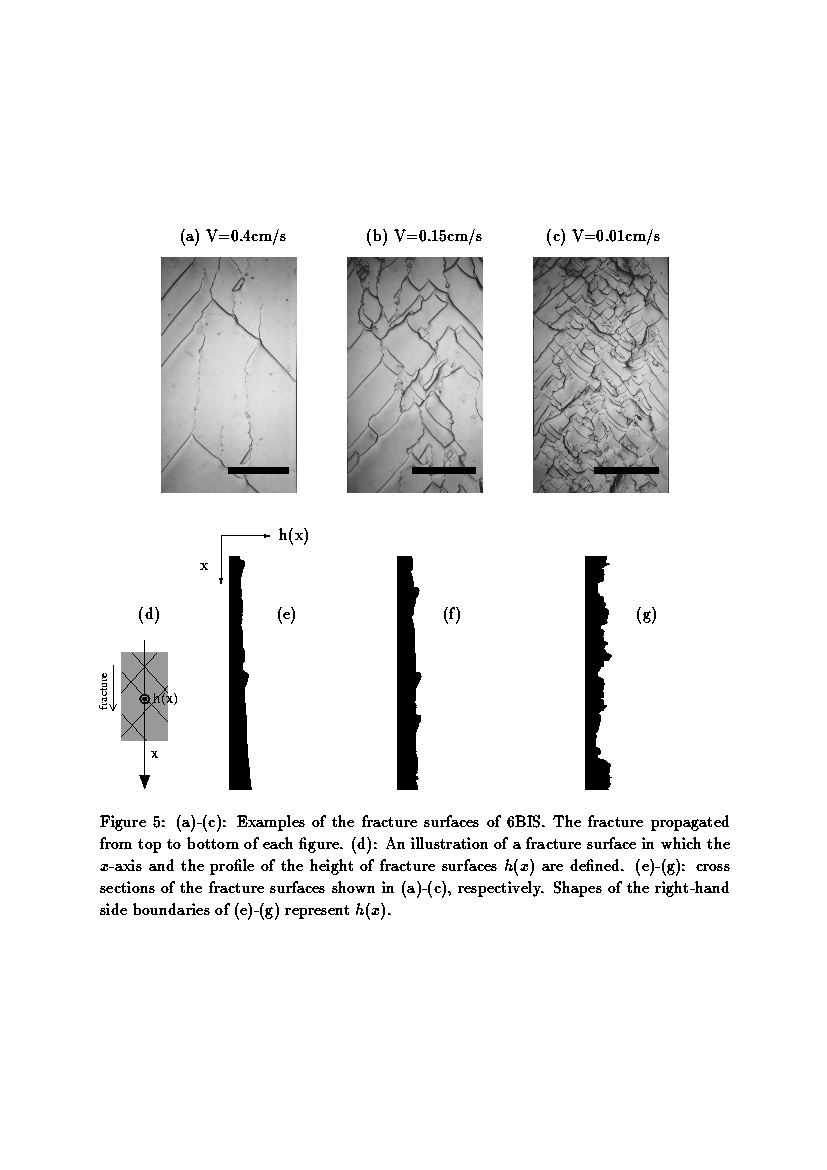) !}

\vspace{4cm}
\end{center}
\caption{(a)-(c): Examples of the fracture surfaces of 6BIS.
The fracture propagated from top to bottom of
each figure. The bars represent 0.9 cm.
(d): An illustration of a fracture surface in which the $x$-axis and
the profile of the height of fracture surfaces $h(x)$ are defined.
(e)-(g): cross sections of
the fracture surfaces shown in (a)-(c), respectively.
Shapes of the right-hand side boundaries of (e)-(g)
represent $h(x)$.}

\label{4b}
\end{figure}




\begin{figure}

\begin{center}
\includegraphics*[width=8.5cm]{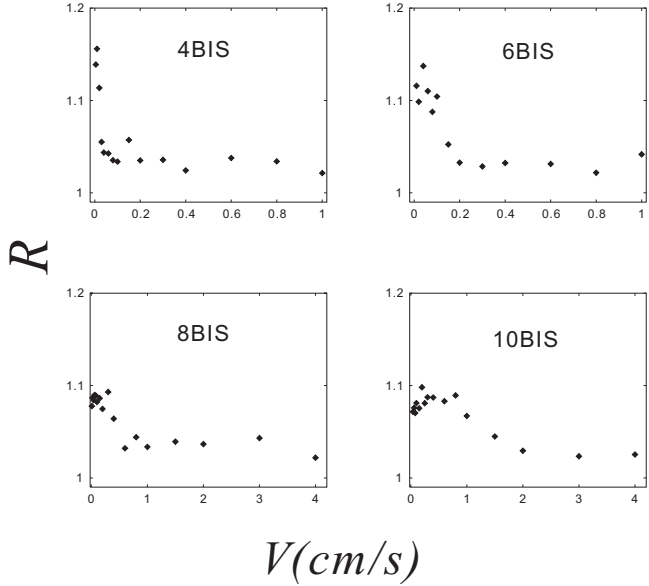}
\end{center}

\caption{An index of the roughness
$ R \equiv \bigl< \sqrt{1+(dh/dx)^2} \ \bigr> $
as a function of $ V $.
As $V$ decreases, $R$ begins to grow up.
The value of $V$ at which the $R$ begin to grow up
(indicated in Fig. 4 by upward arrows)
corresponds to the value of $V$ which
gives the minimum of $G(V)$ in Fig. 4.}
\label{fig:Rough}
\end{figure}

\begin{figure}

\begin{center}
\includegraphics*[width=8cm]{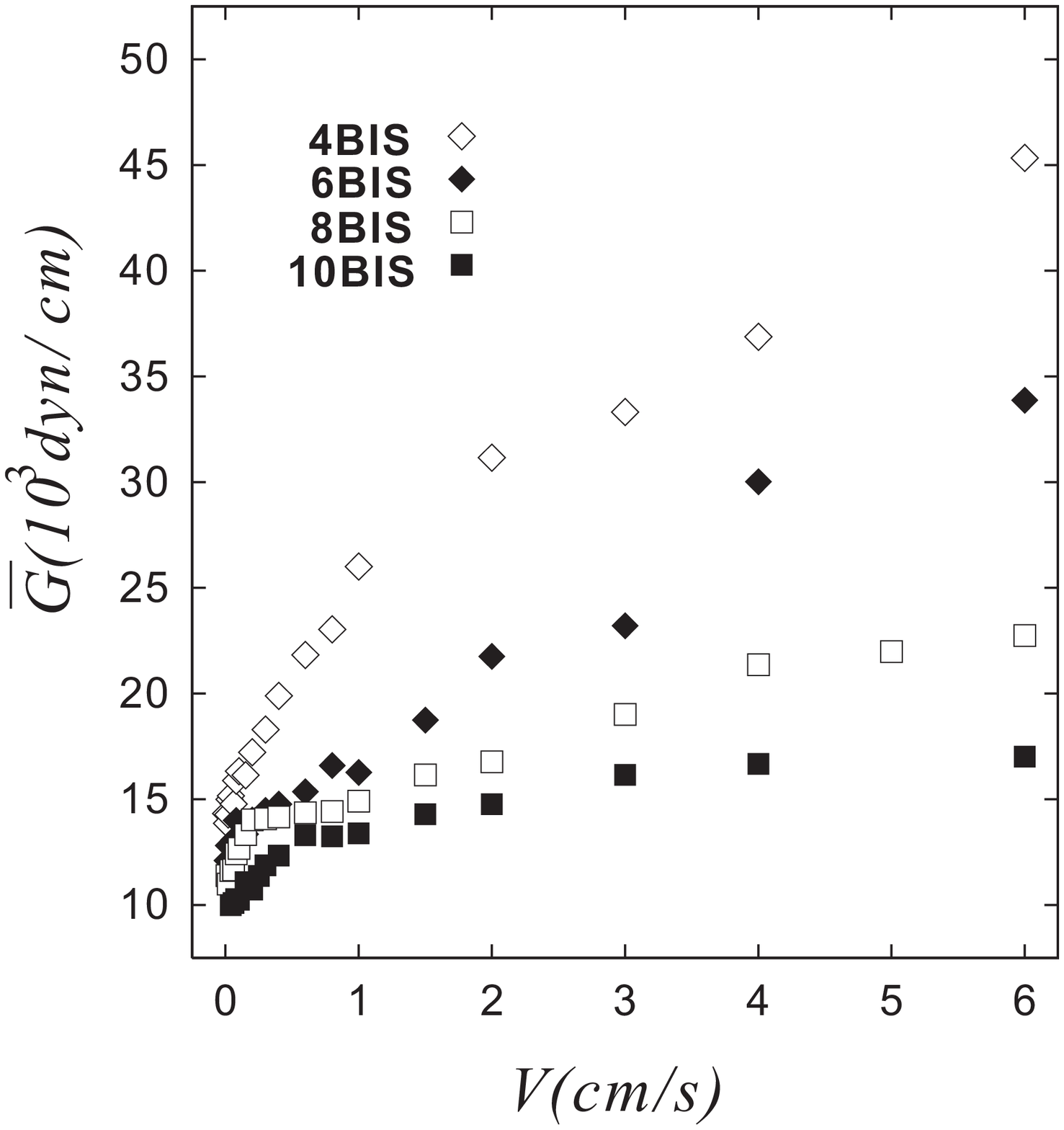}
\end{center}

\caption{The corrected fracture energy $ \overline{G}(V) \equiv G(V)/R(V)^{2}
$.
At fast values of $ V $, behavior of  $ \overline{G}(V) $
is qualitatively identical to that of $G(V)$.}
\label{fig:Rought}
\end{figure}

\begin{figure}

\begin{center}
\includegraphics*[width=7.5cm]{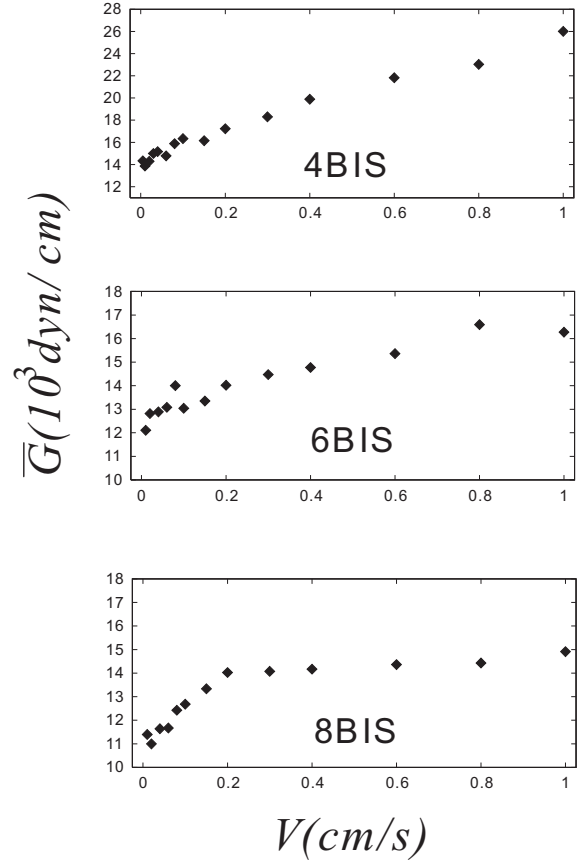}
\end{center}

\caption{$ \overline{G}(V) $ of 4BIS, 6BIS and 8BIS
for $V<$ 1 cm/s. In this region $ \overline{G}(V)$
monotonically depends on $V$. This is quite different from
the behavior of $G(V)$ in Fig. 4.}
\label{fig:cG_v2}
\end{figure}

\end{document}